\documentclass{article}

\usepackage{arxiv_custom}

\usepackage[utf8]{inputenc}
\usepackage[T1]{fontenc}
\usepackage{hyperref}
\usepackage{url}
\usepackage{graphicx}
\usepackage{natbib}
\usepackage{float}
\usepackage{multirow}
\usepackage{amsmath,amssymb,amsfonts}
\usepackage{amsthm}
\usepackage{mathrsfs}
\usepackage[title]{appendix}
\usepackage[HTML]{xcolor}
\usepackage{textcomp}
\usepackage{manyfoot}
\usepackage{booktabs}
\usepackage{algorithm}
\usepackage{algorithmicx}
\usepackage{algpseudocode}
\usepackage{listings}
\usepackage[normalem]{ulem}

\usepackage{pgffor}

\newcommand{\etal}{\textit{et al}.~}
\definecolor{mapColor1}{HTML}{F88379}
\definecolor{mapColor2}{HTML}{FFB570}
\definecolor{mapColor3}{HTML}{FFF275}
\definecolor{mapColor4}{HTML}{A8E6CF}
\definecolor{mapColor5}{HTML}{87CEFA}
\definecolor{mapColor6}{HTML}{CBAACB}

\newcommand{\mapLabelA}[1]{\fcolorbox{white}{mapColor1}{\textcolor{black}{#1}}}
\newcommand{\mapLabelB}[1]{\fcolorbox{white}{mapColor2}{\textcolor{black}{#1}}}
\newcommand{\mapLabelC}[1]{\fcolorbox{white}{mapColor3}{\textcolor{black}{#1}}}
\newcommand{\mapLabelD}[1]{\fcolorbox{white}{mapColor4}{\textcolor{black}{#1}}}

\newcommand{\templidinst}[1]{$\templidinstIn{#1}$}

\newcommand{\templidinstIn}[1]{\big(
    \foreach \x [count=\xi from 1] in {#1} {
      \ifnum\xi>1,\,\fi
      (\x, \x)
    }
\big)}

\title{Saturation-Based Atom Provenance Tracing in Chemical Reaction Networks}

\author{Marcel Friedrichs\\
	Algorithmic Cheminformatics\\
        Faculty of Technology; Center for Biotechnology (CeBiTec), Bielefeld University\\
        Universitätsstraße 25, 33615 Bielefeld, Germany\\
        \texttt{research@mfriedrichs.me}\\
        \texttt{ORCID: \href{https://orcid.org/0000-0001-9846-7212}{0000-0001-9846-7212}}\\
	\And
	Daniel Merkle\\
	Algorithmic Cheminformatics\\
        Faculty of Technology; Center for Biotechnology (CeBiTec), Bielefeld University\\
        Universitätsstraße 25, 33615 Bielefeld, Germany\\
        \texttt{daniel.merkle@uni-bielefeld.de}\\
    Department of Mathematics and Computer Science\\
        University of Southern Denmark\\
        Odense M DK-5230, Denmark\\
        \texttt{daniel@imada.sdu.dk}\\
        \texttt{ORCID: \href{https://orcid.org/0000-0001-7792-375X}{0000-0001-7792-375X}}\\
}

\renewcommand{\shorttitle}{Saturation-Based Atom Provenance Tracing in Chemical Reaction Networks}

\hypersetup{
pdftitle={Saturation-Based Atom Provenance Tracing in Chemical Reaction Networks},
pdfsubject={q-bio.MN, cs.DM},
pdfauthor={Marcel Friedrichs, Daniel Merkle},
}

\begin{document}

\maketitle

\begin{abstract}
Atom tracing is essential for understanding the fate of labeled atoms in biochemical reaction networks, yet existing computational methods either simplify label correlations or suffer from combinatorial explosion. We introduce a saturation-based framework for enumerating labeling patterns that directly operates on atom–atom maps without requiring flux data or experimental measurements. The approach models reaction semantics using Kleisli morphisms in the powerset monad, allowing for compositional propagation of atom provenance through reaction networks. By iteratively saturating all possible educt combinations of reaction rules, the method exhaustively enumerates labeled molecular configurations, including multiplicities and reuse.
Allowing arbitrary initial labeling patterns - including identical or distinct labels - the method expands only isotopomers reachable from these inputs, keeping the configuration space as small as necessary and avoids the full combinatorial growth characteristic of previous approaches. In principle, even every atom could carry a distinct identifier (e.g., tracing all carbon atoms individually), illustrating the generality of the framework beyond practical experimental limitations.
The resulting template instance hypergraph captures the complete flow of atoms between compounds and supports projections tailored to experimental targets. Customizable labeling sets significantly reduce generated network sizes, providing efficient and exact atom traces focused on specific compounds or available isotopes. Applications to the tricarboxylic acid cycle, and glycolytic pathways demonstrate that the method fully automatically reproduces known labeling patterns and discovers steady-state labeling behavior. The framework offers a scalable, mechanistically transparent, and generalizable foundation for isotopomer modeling and experiment design.
\end{abstract}



\section{Introduction}\label{sec1}

Atom tracing has become a fundamental tool in biochemical research, enabling researchers to study metabolic pathways and chemical transformations. This is typically achieved by labeling one or more atoms, such as \( {}^{13}\mathbf{C} \) carbon atoms, in metabolites or compounds. By introducing labeled species into a chemical or metabolic system, researchers can track their fate in downstream compounds using analytical techniques such as nuclear magnetic resonance (NMR) or mass spectrometry (MS)~\citep{Jang2018}. These measurements are essential for applications including pathway elucidation, metabolic flux analysis, and knockout experiments.

However, planning and executing atom tracing experiments is both costly and time-intensive, owing to specialized tracer synthesis, sample preparation, and high-resolution MS/NMR analysis~\citep{Jang2018, deFalco2022}. Identifying the most informative labeling patterns is crucial for maximizing the insights gained, while minimizing the number of labeled atoms can significantly reduce experimental costs. Consequently, computational methods that support the design and analysis of labeling strategies have become increasingly valuable~\citep{Bey2021}.

Computational cheminformatics approaches address this need by enabling systematic enumeration and evaluation of labeling patterns \emph{in silico}, thereby guiding and refining experimental design. Several influential computational frameworks have been proposed to model isotope distributions in chemical reaction networks (CRNs):
\begin{itemize}
    \item \emph{Isotopomer networks}, introduced by \cite{Schmidt1997}, represent labeled compounds using binary vectors that denote the presence or absence of a label at each atomic position. While conceptually straightforward, this leads to a combinatorial explosion, each compound with $N$ labelable atoms yields $2^N$ isotopomers, making the method impractical for large systems. Additionally, the created isotopomer mapping matrices (IMM) cannot capture multiplicities (i.e., distinct labeling patterns of the same compound in a single reaction) or different label types.

    \item \cite{Wiechert1999} expand the isotopomer network method by using cumulative subsets (called \emph{cumomers}) that represent the probability that a specific subset of carbon positions is labeled, regardless of the labeling of other positions. This reduces and simplifies the state space and equations generated in the network.

    \item \cite{Antoniewicz2007} took previous isotopomer network approaches and developed a bottom-up method called \emph{elementary metabolite units} (EMU) which drastically reduces the state space. In short, measured fragment labelings from, for example, NMR in desired target compounds are tracked backwards through the reactions, eliminating not reachable labelings.

    \item \cite{Njgaard2021} developed a new method based on \emph{transformation semigroups}. Atoms are traced through a CRN by applying transformations representing reactions. A Cayley graph is constructed representing atom states as vertices and transformations as edges. Similar to EMUs, a projected Caley graph can be constructed immediately reducing the graph to atoms of interest and, therefore, increasing performance and making large networks feasible. However, it is impossible to represent multiplicities and transformations can represent only parts of a reaction.

    \item \cite{Golnik2025} proposed atom transition networks that track how a single labeled atom propagates through a chemical reaction network using atom-atom maps. The solution gives the steady-state distribution of label probabilities across atoms in the network.
\end{itemize}
Despite their differences, all these approaches rely on accurate atom-atom mappings for chemical reactions. Such mappings can be obtained experimentally, sourced from curated databases, computed using rule-based graph rewriting systems~\citep{Andersen2016}, or reaction mechanism generators such as RMG~\citep{RMG3}. Isotopomer and EMU methods often require additional data, such as flux measurements or known fragment labelings, which may not be readily available.

In this work, we introduce a new approach to isotopomer modeling that captures full atom provenance, including multiplicities and reuse, and relies solely on atom-atom maps as input. We formalize the semantics of this method using a \emph{Kleisli morphism} operating on a \emph{monadic structure}, enabling the seamless propagation of atom provenance through reaction networks, providing a scalable and expressive framework for atom tracing in complex CRNs. Applied to well-studied examples, such as the tricarboxylic acid cycle and glycolysis pathways, we demonstrate the methods capability to fully automatically reproduce known labeling patterns from published analyses~\citep{Dong2019, MIT130611, Njgaard2021, Andersen2014} and discover steady-state labeling behavior. The richness of atom provenance produced for these labeling patterns was previously only possible by manual atom tracing. The thesis by~\citep{Dong2019} demonstrates this in analyzing metabolic pathways in the context of cancer. Based on published and experimental results, all atom labeling patterns were meticulously but manually traced. For the tricarboxylic acid cycle (TCA), this meant manually tracing all cycle iterations for all possible labeling patterns, which is labor-intensive and the possibility of making mistakes is high. In contrast, our new method is now capable of performing such traces fully automatic for all labeling patterns that arise from all cycle iterations.


\section{Methods}\label{sec2}

We present a saturation-based atom-tracing framework that constructs the reachable labeling configuration space of a chemical reaction network by iteratively applying chemical reactions. Each reaction is encoded as a list of educts and products with a slot-to-position mapping, and label propagation is formalized through Kleisli composition within a monadic structure. This approach supports exact atom-level provenance, multiplicities, reuse, and combinatorial educt pairings, enabling exhaustive enumeration of reachable labeled molecular configurations.

Unlike isotopomer network or EMU-based models that allow only binary labeling patterns, our approach preserves full atom-level and labeling dependencies across successive reaction steps. By embedding reaction semantics into a compositional categorical model, the framework unifies biochemical label tracing and categorical reasoning - providing a rigorous algebraic foundation for atom provenance.

This methods section strikes a balance between formal definition and practical implications and closely follows the reference implementation in terms of minimal input requirements. While atom–atom maps are fundamentally derived from the molecular graph structure of educts and products—potentially including stereochemical constraints—the execution of the method operates exclusively on these maps. As a consequence, compounds are reduced to sets of unique position IDs, since all structural information relevant for atom tracing is already encoded in the atom–atom mappings supplied for each reaction. This separation highlights the advantage of graph-grammar–based approaches, where atom–atom maps arise naturally as a consequence of rule application rather than as independent annotations. A separate formal definition of the proposed method is available in \autoref{sec:appendix}.

\begin{figure}[H]
    \centering
    \includegraphics[width=\textwidth]{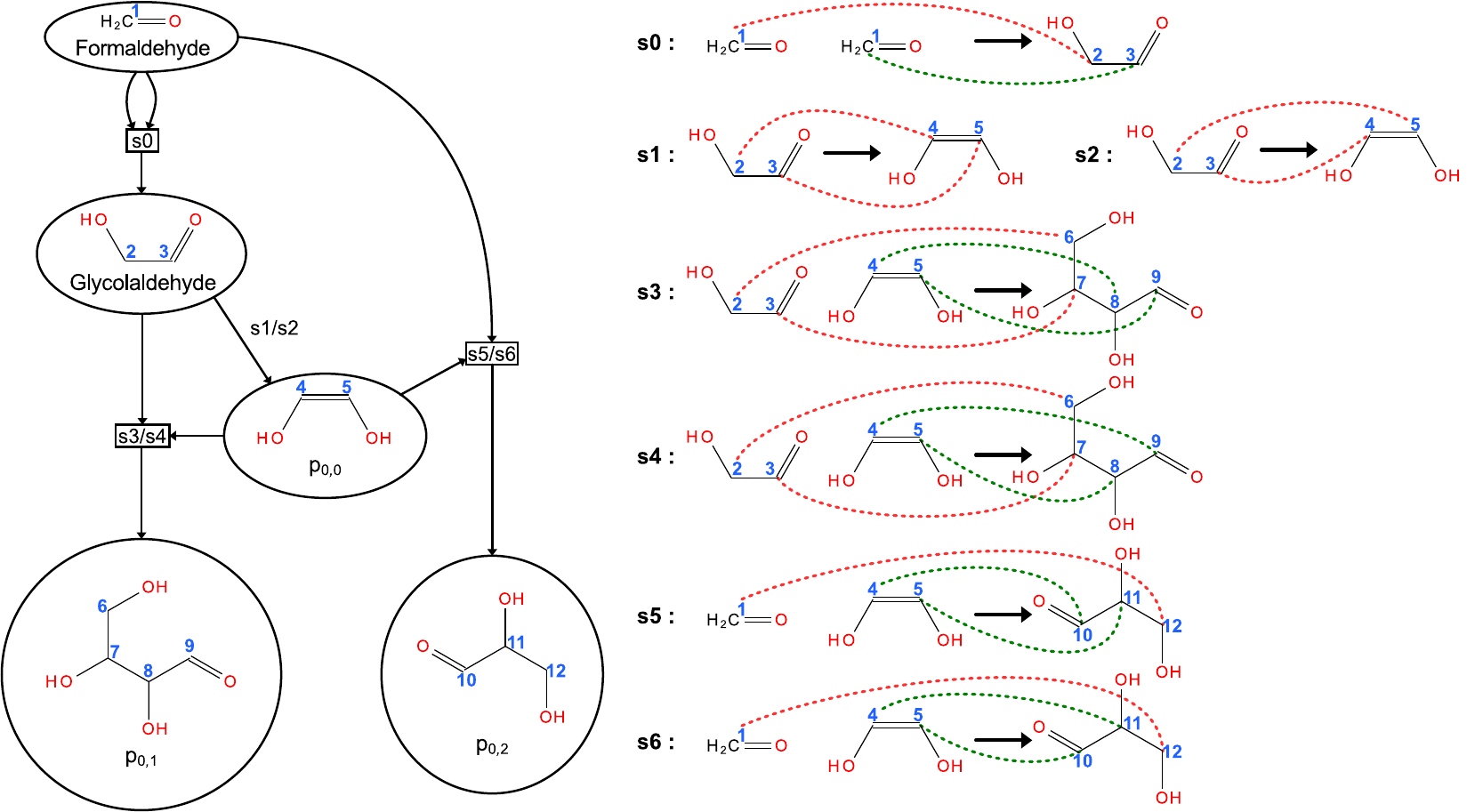}
    \caption{Left: Chemical reaction network CRN hypergraph of a formaldehyde toy network generated by a graph rewriting system from a set of reaction rules. Carbon atoms of all compounds are assigned unique position IDs in blue. Right: Carbon atom-atom maps of all reactions and symmetries labeled $s0$ to $s6$. Three pairs are symmetries of each other: $(s1, s2)$, $(s3, s4)$, and $(s5, s6)$.}
    \label{fig:formaldehyde_dg}
\end{figure}

First, we define the core elements of our saturation-based approach. In a chemical \emph{reaction network} compounds participate in chemical reactions, producing defined product compounds from educt compounds. These compounds in the network can be assigned unique IDs to their individual atoms, subsequently called a \emph{compound template}. The template name follows the analogy that, in a real chemical system, there exist multiple molecules/\emph{instances} of the same compound species sharing the same set of assigned atom IDs. For tracing atoms through the reaction network, \emph{labeled template instances} extend a \emph{compound template} by assigning each uniquely identified atom with an origin. This origin represents from which \emph{compound template} this atom came from before traversing the reaction network and ending up at this specific position in this specific \emph{labeled template instance}. There exists exactly one compound template for each individual compound in a reaction network, but multiple labeled template instances can exist for each individual compound due to varying origin information. \emph{Reaction rules} consist of a list of educt compound templates, a list of product compound templates, and an exact mapping between the atoms of the two lists. When a reaction rule is applied, a list of labeled template instances with multiplicities and reuse are chosen and used to produce product labeled template instances.
Second, definitions for the monad and Kleisli composition enabling the saturation process are provided.
Third, the concurrent construction of a template instance graph and possible graph projections are described.

\subsection{Chemical Reaction Networks and Atom Maps}

A prerequisite for this method is the existence of a chemical reaction network consisting of chemical compounds and reactions, where each reaction is associated with one or more admissible atom–atom maps. A single chemical reaction may therefore give rise to multiple distinct atom–atom mappings, for example due to molecular symmetries. Throughout this work we treat atom–atom maps as part of the input: not every graph automorphism of a molecular representation corresponds to a different bijection between educt and product atoms, since stereospecificity and other mechanistic constraints can rule out certain mappings. These atom-atom maps provide the means of our approach to trace the flow of atoms through the reaction network. An example formaldehyde reaction network with atom-atom maps used in subsequent definitions is visualized in \autoref{fig:formaldehyde_dg}. The approach is generic for whole compounds and all atom types. For brevity, only carbon atoms are considered in this example. Three reaction atom-atom map pairs are symmetries of each other: $(s1, s2)$, $(s3, s4)$, and $(s5, s6)$.

\subsection{Compound Templates}

\begin{figure}[H]
    \centering
    \includegraphics[width=0.75\textwidth]{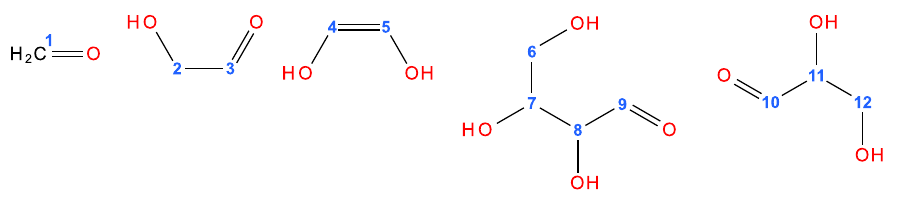}
    \caption{Unique carbon atom position IDs in blue for formaldehyde, glycolaldehyde, 1,2-ethendiol, 	
2,3,4-trihydroxybutanal, and glyceraldehyde. Their representation as compound templates is $(1)$, $(2, 3)$, $(4, 5)$, $(6, 7, 8, 9)$, and $(10, 11, 12)$.}
    \label{fig:example_C_labelings}
\end{figure}

To uniquely identify individual atoms of all compounds in the CRN, let $\Omega \subset \mathbb{Z}$ be the global set of linearized position/atom IDs. The example visualized in \autoref{fig:example_C_labelings} defines 12 position IDs $\{1, 2, \dots, 11, 12\}$ regarding only carbon atoms of all compounds.

Atoms are grouped together as compounds which is not represented by $\Omega$.
For this, let a \emph{compound template} $T$ be a totally ordered set $T=\{p_1,\dots,p_{|T|}\}\subseteq \Omega$ of position IDs with the binary relation $\leq$. Position IDs are unique to one template: $\forall (T_a, T_b) \in \text{CRN} \times \text{CRN}, T_a \ne T_b: T_a \cap T_b = \emptyset$. A compound template may represent all atoms of a compound or only a subset, for example all carbon atoms, depending on the use-case. However, it would also be feasible to regard the union of multiple compounds as a compound template. The five compounds used in the reaction network of \autoref{fig:formaldehyde_dg} with assigned position IDs to carbon atoms are visualized in \autoref{fig:example_C_labelings}.

\subsection{Labeled Template Instances}

In labeling experiments, specific atoms are labeled with isotopes such as $^{13}C$. As the labeled compound still represents the unlabeled compound in structure and function, it can be regarded as a specific labeled instance of that compound. Following the assignment of position IDs to the atoms of that compound as a compound template, the labeled instances can, therefore, be regarded as a \emph{labeled template instance}.
\newpage
Formally, a labeled template instance $\widehat{T}$ is a tuple: $\widehat{T}=\Big((o_i, p_i)~\big|~i \in 1, \dots|T|\Big)$ where:
\begin{itemize}
    \item $o_i$ is an arbitrary \emph{origin position ID} from $\Omega$, currently associated with position ID $p_i$. It represents the origin from which compound template this atom came from before traversing the reaction network and ending up at this specific position in this specific labeled template instance, i.e. the atom provenance.
    \item $p_i$ is a specific \emph{position ID} of template $T$.
\end{itemize}
Two labeled template instances of the same compound template are considered distinct whenever they differ in at least one origin position ID while sharing the same position IDs. For example, the labeled template instance $\big((1, 2), (1, 3)\big)$ corresponds to glycolaldehyde, where both carbon atoms at positions $2$ and $3$ originate from atoms with position ID $1$ in (two copies of) formaldehyde. In contrast, $\big((2, 2), (3, 3)\big)$ is the identity template instance of glycolaldehyde and represents the unlabeled starting molecule, where each carbon atom originates from itself. Labeled template instances can represent arbitrary labeling patterns arising from the combinatorics of origin position IDs, including all realistic isotopomer patterns used in practice. \emph{Identity template instances} $\widehat{T}_{id}$ have atom provenance only from themselves with equal origin and position IDs, e.g. $\forall i \in 1, \dots, |T| : o_i=p_i$. Considering the compounds of \autoref{fig:example_C_labelings}, the identity template instances would be 
\templidinst{1}, \templidinst{2, 3}, \templidinst{4, 5}, \templidinst{6, 7, 8, 9}, and \templidinst{10, 11, 12}.

\subsection{Configurations}\label{sec2_configurations}

For chemical reactions to occur, one or more labeled template instances need to be defined. This set of instances, subsequently called a \emph{configuration}, is the single source of educts for the reaction rules of the network. A configuration defines the starting point of the presented method and is iteratively expanded in the saturation process.

Let $\mathbf{Inst}$ denote the set of all possible labeled template instances. Chemically, this represents the unrealistic existence of $|\Omega|$ unique labels, such as $^{13}C$, and that all compounds of the reaction network exist with all possible combinations of labeled atoms. This, however, quickly leads to a combinatorial explosion, which is why this method focuses on subsets of $\mathbf{Inst}$ as configurations. Let $\mathcal{X} = \mathcal{P}_{fin}(\mathbf{Inst})$ be the set of all finite subsets of $\mathbf{Inst}$. A \emph{configuration} $X \in \mathcal{X}$ is then a finite set of labeled template instances. The \emph{initial configuration} for a chemical reaction network CRN, from which the saturation process begins, is comprised only of identity template instances:
\begin{align*}
X_0 = \Big\{\widehat{T}_{id} \mid T \in \text{CRN}\Big\}
\end{align*}
Using again the example of \autoref{fig:example_C_labelings}, the set of all template instances $\mathbf{Inst}$, i.e. the full combinatorics, allows for $12^1 + 2*12^2 + 12^3 + 12^4 = 22,764$ different labeled template instances. For readability, the abbreviated set would be
\begin{align*}
    \Big\{&\big((1, 1)\big), \big((2, 1)\big), \dots, \big((12, 1)\big), \big((1, 2), (1, 3)\big), \big((1, 2), (2, 3)\big), \dots, \big((12, 2), (12, 3)\big),\\
    &\dots, \big((12, 10), (12, 11), (12, 12)\big)\Big\}
\end{align*}
and the initial configuration $X_0$ would be
\begin{align*}
    \Big\{\templidinstIn{1}, \templidinstIn{2, 3}, \templidinstIn{4, 5}, \templidinstIn{6, 7, 8, 9}, \templidinstIn{10, 11, 12}\Big\}.
\end{align*}

\begin{figure}[H]
    \centering
    \includegraphics[width=0.75\textwidth]{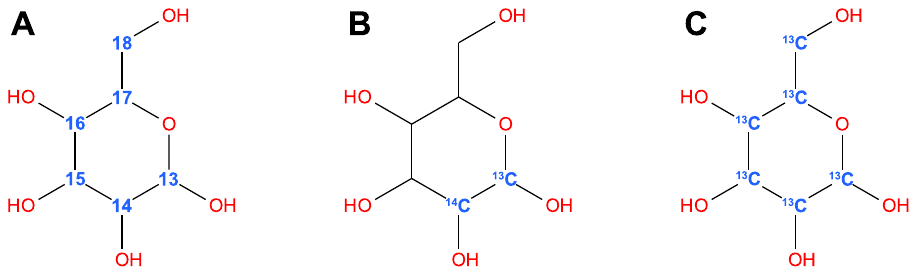}
    \caption{A: Glucose with position IDs assigned to carbon atoms in blue. \mbox{B: Glucose-1-$^{13}C$-2-$^{14}C$}. \mbox{C: Glucose-U-$^{13}C_6$}.}
    \label{fig:example_labeling_glucose}
\end{figure}

The initial configuration $X_0$ will consider all atom position IDs as uniquely labeled and result in a maximal saturation for CRN. Large networks consisting of many compounds can in turn result in a large number of generated labeled template instances. For most practical scenarios, however, only a single compound $T^v$ is of particular interest. The initial configuration can be chosen, so that only $T^v$ has all atoms uniquely labeled and all other compound's atoms are considered equal, e.g. unlabeled. This significantly reduces the number of generated labeled template instances to a maximum of $\sum_{T \in \text{CRN}} (|T^v| + 1)^{|T|}$. Initial template instances except $T^v$ are considered unlabeled which can be denoted by setting all origin position IDs to the same arbitrary ID not used by any numbered atom. Here, the position ID $-1$ is used and, for readability, this unlabeled position ID is visualized as $\bigcirc$.
\begin{align*}
X_{0,T^v} = \Big\{\big((\bigcirc, p_1), \dots, (\bigcirc, p_{|T|})\big) \mid T \in \text{CRN} \setminus \{T^v\}\Big\} \bigcup \Big\{\widehat{T}^v_{id}\Big\}
\end{align*}
This performance optimization can be expanded further by considering only specific atom origins. Consider the Glucose with assigned carbon position IDs from \autoref{fig:example_labeling_glucose}A. The specific \mbox{Glucose-1-$^{13}C$-2-$^{14}C$}  labeling of \autoref{fig:example_labeling_glucose}B would result in the template instance $\big((13, 13), (14, 14), (\bigcirc, 15), (\bigcirc, 16), (\bigcirc, 17), (\bigcirc, 18)\big)$. To represent specific labeling experiments where atoms are not differentially labeled, such as fully labeled \mbox{Glucose-U-$^{13}C_6$} of \autoref{fig:example_labeling_glucose}C, the initial configuration can be further modified. As done before with the unlabeled position ID, other origin position IDs can be reused as well, not specifically representing a single position ID anymore, but rather a set of position IDs. The fully $^{13}C$ labeled Glucose, for example, can be used in the initial configuration as template instance $\big((13, 13), (13, 14), (13, 15), (13, 16), (13, 17), (13, 18)\big)$ where an origin ID $13$ now represents any carbon of Glucose. This further reduces the number of generated labelings. For simplicity, the origin ID representing such label sets is always chosen to be the minimum position ID of that set, in this case $13$.

\subsection{Reaction Rules}

Let $\mathcal{R}$ denote the finite set of all reaction rules in CRN. A reaction rule $R \in \mathcal{R}$ is a pair of educts and products $(R_E, R_P)$ where:
\begin{itemize}
    \item $R_P = (P_1, \dots, P_m)$ is the ordered list of product templates with product position IDs.
    \item $R_S = \sum_{P \in R_P} |P|$ is the number of required slots to identify all position IDs of all products.
    \item $R_E = (E_1, \dots, E_k)$ is the ordered list of educt templates with slot mappings: Each $E_i$ is a map $c_{E_i}: p \rightarrow s~|~p \in T_{E_i}, s \in [0,R_S)$ assigning each position ID in the template $T_{E_i}$ to a slot index $s$.
\end{itemize}
The rule induces a total mapping $c_R: \{0, \dots, R_S - 1\} \rightarrow \text{\textit{product position IDs}}$ from educt position IDs to product position IDs, determined by the ordered product list.

\begin{figure}[H]
    \centering
    \includegraphics[width=\textwidth]{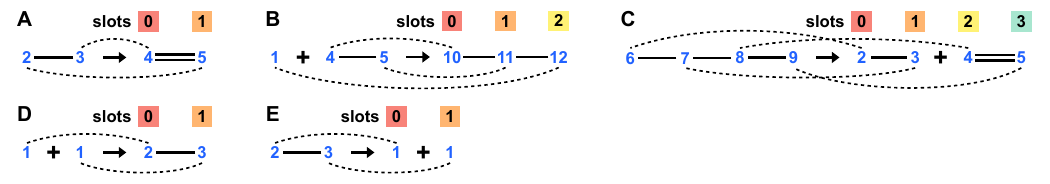}
    \caption{Five reaction rules from the formaldehyde toy reaction network showing the mapping of educt to product position IDs via slot indices. The slot indices are shown above the product position IDs. Dashed lines represent the mapping from educt position IDs to slot indices. A: $s2$, B: $s5$, C: $reverse(s3)$, D: $s0$, and E: $reverse(s0)$.}
    \label{fig:example_reaction_rule_depictions}
\end{figure}

\noindent Several reaction rules from \autoref{fig:formaldehyde_dg} and reverse reactions can be defined as follows and as visualized in \autoref{fig:example_reaction_rule_depictions} using a total mapping from educt to product position IDs via slot indices:
\begin{itemize}
    \item \emph{Single educt to single product} (\autoref{fig:example_reaction_rule_depictions}A):\\
        $s2 = \Big(\big(\{2 \rightarrow \mapLabelB{1},~3 \rightarrow \mapLabelA{0}\}\big), \big((\mapLabelA{4}, \mapLabelB{5})\big)\Big)$
    \item \emph{Multiple educts to single product} (\autoref{fig:example_reaction_rule_depictions}B):\\
        $s5 = \Big(\big(\{1 \rightarrow \mapLabelC{2}\},~\{4 \rightarrow \mapLabelA{0},~5 \rightarrow \mapLabelB{1}\}\big), \big((\mapLabelA{10}, \mapLabelB{11}, \mapLabelC{12})\big)\Big)$
    \item \emph{Single educt to multiple products} (\autoref{fig:example_reaction_rule_depictions}C):\\
        $reverse(s3) = \Big(\big(\{6 \rightarrow \mapLabelA{0},~7 \rightarrow \mapLabelB{1},~8 \rightarrow \mapLabelC{2},~9 \rightarrow \mapLabelD{3}\}\big), \big((\mapLabelA{2}, \mapLabelB{3}),~(\mapLabelC{4}, \mapLabelD{5})\big)\Big)$
    \item \emph{Educt multiplicity} (\autoref{fig:example_reaction_rule_depictions}D):\\
        $s0 = \Big(\big(\{1 \rightarrow \mapLabelA{0}\},~\{1 \rightarrow \mapLabelB{1}\}\big), \big((\mapLabelA{2}, \mapLabelB{3})\big)\Big)$
    \item \emph{Product multiplicity} (\autoref{fig:example_reaction_rule_depictions}E):\\
        $reverse(s0) = \Big(\big(\{2 \rightarrow \mapLabelA{0},~3 \rightarrow \mapLabelB{1}\}\big), \big((\mapLabelA{1}),~(\mapLabelB{1})\big)\Big)$
\end{itemize}

\subsection{Rule Application With Reuse}

For a reaction rule to be applied, a list of educt labeled template instances needs to be chosen to participate in that reaction. As the reaction rule defines the compound templates $R_E$, only labeled template instances of those templates are valid choices. Given a configuration $X \in \mathcal{X}$ and a reaction rule $R$, an educt combination $c_R$ of $R$ is defined as a mapping from all educt templates $R_E$ to labeled template instances in $X$ of the same template. The same template instance may be chosen for multiple educts (reuse). For each educt combination:
\begin{itemize}
    \item For each slot index $s \in R_S$, take the origin ID from the corresponding atom.
    \item Assign the product position ID $c_R(s)$ to that origin ID.
    \item Group the resulting labeled atoms into product template instances in order.
\end{itemize}
The output is a new configuration $X' \in \mathcal{X}$ containing the resulting product template instances.

As an example, consider a configuration of three template instances a, b, and c, where the origin of position ID $1$ for b is another position ID not shown in \autoref{fig:formaldehyde_dg}:
\begin{align*}
    X = \Big\{\underbrace{\big((1, 1)\big)}_a, \underbrace{\big((23, 1)\big)}_b, \underbrace{\big((4, 4), (5, 5)\big)}_c\Big\}
\end{align*}
The reaction rule $s0$ can be applied with four different educt lists from $X$, each resulting in a different following configuration, where two labeled template instances of the same template $(1)$ as educts are combined into a single new labeled template instance of a different template $(2, 3)$ as product:
\begin{align*}
    &(a, a): X'_1 = X \cup \Big\{\big((1, 2), (1, 3)\big)\Big\}\\
    &(a, b): X'_2 = X \cup \Big\{\big((1, 2), (23, 3)\big)\Big\}\\
    &(b, a): X'_3 = X \cup \Big\{\big((23, 2), (1, 3)\big)\Big\}\\
    &(b, b): X'_4 = X \cup \Big\{\big((23, 2), (23, 3)\big)\Big\}
\end{align*}
Similarly, the reaction rule $s6$ can be applied, resulting in two different following configurations:
\begin{align*}
    &(a, c): X'_5 = X \cup \Big\{\big((5, 10), (4, 11), (1, 12)\big)\Big\}\\
    &(b, c): X'_6 = X \cup \Big\{\big((5, 10), (4, 11), (23, 12)\big)\Big\}
\end{align*}

\subsection{Free Monoid, Powerset Monad, and Kleisli Category}

Applying reaction rules requires the enumeration of all educt combinations from a configuration set of template instances $X \in \mathcal{X}$ with reuse. For example, the reactions $s1$ and $s2$ of the formaldehyde CRN could be applied to different glycolaldehyde template instances, such as $((1, 2), (1, 3))$ or $((2, 2), (3, 3))$. The set of all finite lists of elements of $X$ is defined as
\begin{align*}
    X^* = \bigcup\limits_{n \geq 0} X^{\times n},
\end{align*}
where $X^{\times 0} = \{\epsilon\}$ consists of the empty list. The free monoid on $X$ is the monoid $(X^*, \cdot, \epsilon)$, where the binary operation $\cdot : X^* \times X^* \rightarrow X^*$ is concatenation of lists, and $\epsilon$ is the empty list. This structure provides the multiplicity, reuse, and order required as described before.

Given a configuration $X \in \mathcal{X}$ the application of each educt combination of each rule produces a new configuration $X' \in \mathcal{X}$. For this, the powerset monad is defined as the triple $(\mathcal{P}_{fin}, \eta, \mu)$ consisting of the endofunctor $\mathcal{P}_{fin} : \mathbf{Set} \rightarrow \mathbf{Set}$ and the two natural transformations $\eta$ and $\mu$. The unit $\eta$ is defined as $\eta_Y : Y \rightarrow \mathcal{P}_{fin}(Y)$, assigning to an element of $Y$ the set consisting of that single element $\eta_Y(y) = \{y\}$. The multiplication $\mu$ flattens nested finite sets into a single set through union: $\mu_Y : \mathcal{P}_{fin}(\mathcal{P}_{fin}(Y)) \rightarrow \mathcal{P}_{fin}(Y)$, $\mu_Y(y) = \bigcup y$.

Further, the Kleisli category $C_{\mathcal{P}_{fin}}$ for the powerset monad is defined, with sets as objects and morphisms $f : Y \rightarrow Z$ of the form $Y \rightarrow \mathcal{P}_{fin}(Z)$. Given two morphisms $f : Y \rightarrow \mathcal{P}_{fin}(Z)$ and $g : Z \rightarrow \mathcal{P}_{fin}(W)$ their Kleisli composition is defined as $g \odot f = \mu_W \circ \mathcal{P}_{fin}(g) \circ f$.

Given a configuration $X \in \mathcal{X}$ the rule application morphism \emph{step} is defined as $step : \mathcal{P}_{fin}(Y) \rightarrow \mathcal{P}_{fin}(\mathcal{P}_{fin}(Y))$ in $C_{\mathcal{P}_{fin}}$ which applies all rules $R \in \mathcal{R}$ on all educt combinations $c_R$ of $X^*$ and produces a set of new configurations by:
\begin{align*}
    step(X) = \big\{ X \cup R_P(c_R) \mid R \in \mathcal{R}, c_R \in X^* \big\}
\end{align*}
Because $\mathcal{R}$ and $c_R$ are both finite, $step(X)$ is a finite set of successor configurations. Flattening via $\mu$, $\mu_Y \circ step : \mathcal{P}_{fin}(Y) \rightarrow \mathcal{P}_{fin}(Y)$ can be viewed as a Kleisli morphism $step : Y \rightarrow_{C_{\mathcal{P}_{fin}}} Y$, representing a single transition step of the system. Given the formaldehyde example from \autoref{fig:formaldehyde_dg} and a configuration with only one formaldehyde template instance $((1, 1))$, $step$ finds all possible reactions and educt combinations that can be applied. In this case, it would only be reaction $s0$ resulting in a single new configuration consisting of the original formaldehyde template instance and the new glycolaldehyde template instance $((1, 2), (1, 3))$.

\subsection{Saturation Process}

The process of applying reaction rules to a configuration involves iterating through educt combinations and updating the configuration accordingly. Iteratively applying reaction rules expands the configuration with new template instances until a fixed point is reached where no further instances appear. The n-fold Kleisli composition is expressed as
\begin{align*}
    step^{(n)} = \underbrace{step \odot \cdots \odot step}_{n~\text{times}},
\end{align*}
representing all configurations reachable in $n$ steps. A saturation process can be defined by choosing $n$ so that the produced configuration is closed
\begin{align*}
    step^* = \bigcup\limits_{n \geq 0} step^{(n)},
\end{align*}
representing all reachable configurations under repeated application of all rules $\mathcal{R}$ in CRN.

\subsection{Template Instance Hypergraph Construction}

In addition to determining which template instances can be generated using the saturation, tracing the flow of atoms from instance to instance is of equal interest. For this, a directed hypergraph $G = (V, E)$ is generated from the full saturation, where $V$ is the set of vertices and $E$ is the set of directed edges between vertices. The functions $s(e)$ and $t(e)$ for an edge $e \in E$ represent the finite sets of source and target vertices respectively. All template instances are added as vertices in $G$. When a reaction rule is applied on a list of educt template instances, a hyperedge is created in $G$ between those educt and the product template instances. The direction of the hyperedge is defined from the educt to the product instances.

\begin{figure}[H]
    \centering
    \includegraphics[width=\textwidth]{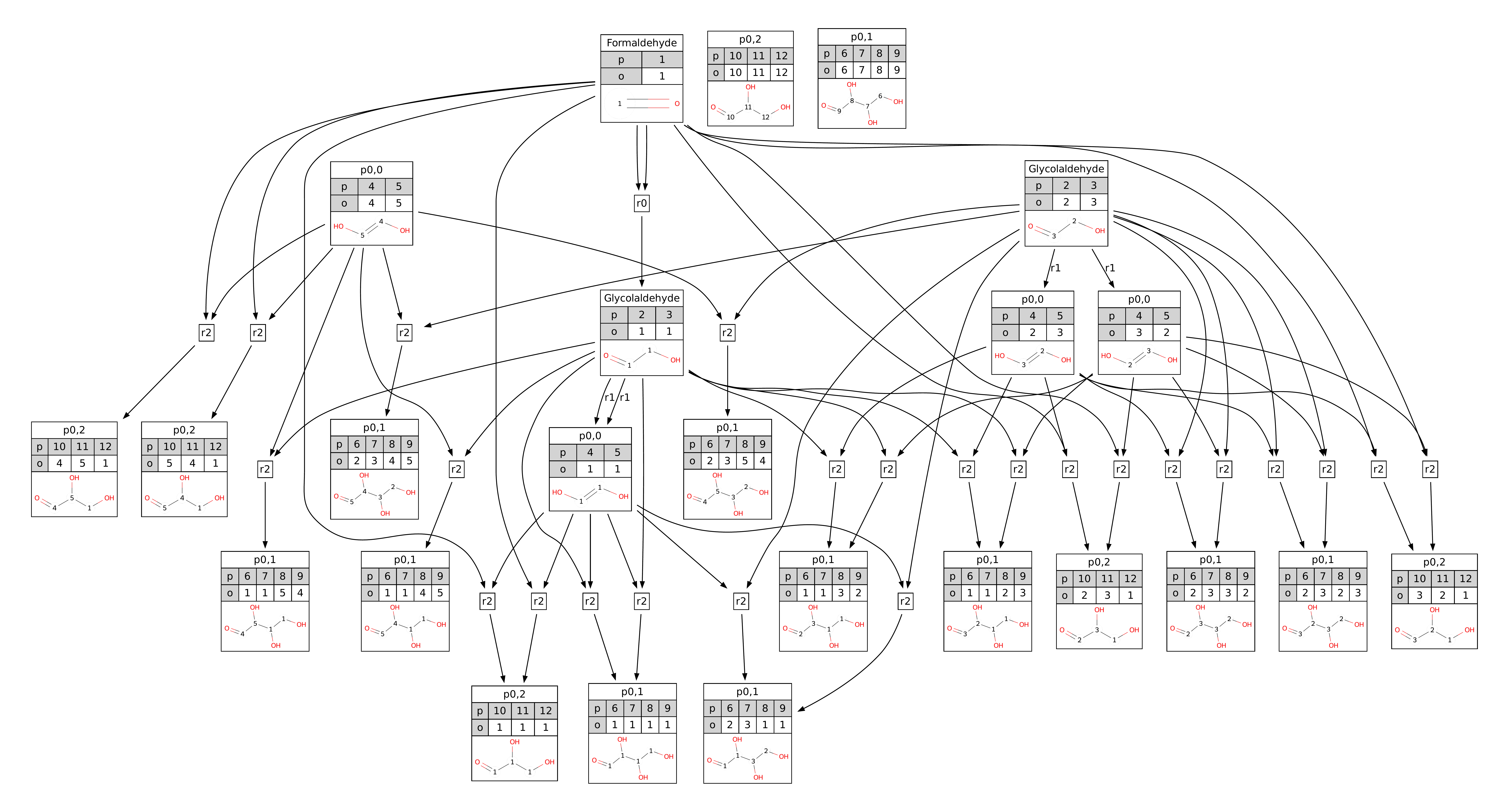}
    \caption{Full saturation graph for the formaldehyde toy network. Each vertex represents a labeled template instance as a table of the compounds name, position and origin IDs, as well as the chemical structure with origin IDs assigned to the carbons. Edge labels represent the reactions that produced this template instance.}
    \label{fig:toy_network_formaldehyde_full_graph}
\end{figure}

Using again the example reaction network of \autoref{fig:formaldehyde_dg}, the hypergraph $G$ after full saturation is visualized in \autoref{fig:toy_network_formaldehyde_full_graph}. It consists of 24 nodes and 29 hyperedges. Of note is reaction rule $s0$ which uses two instances of formaldehyde as educts to produce one glycolaldehyde. This multiplicity with reuse is fully covered by the saturation-based method using all combinations of available formaldehyde template instances in a configuration.

\subsection{Graph Projections}

$G$ represents the full saturation, however, experiments usually focus on a single or only few labeled compounds to be cost-effective and feasible. Let $source$ be a set of position IDs that represent the labeled compounds of interest. Let $V_{source} \subseteq V$ be a vertex set where
\begin{align*}
    V_{source} &= V_{primary} \cup V_{secondary}\\
    V_{primary} &= \{v \in V \mid \{v_{o_1}, \dots, v_{o_{|v|}}\} \cap source \neq \emptyset\}\\
    V_{secondary} &= \{v \in V \mid \exists e \in E, (v \in s(e) \lor v \in t(e)) \land s(e) \cap V_{primary} \neq \emptyset\} 
\end{align*} and $E_{source} \subseteq E$ an edge set where
\begin{align*}
    E_{source} = \{e \in E \mid s(e) \cap V_{source} \neq \emptyset \land t(e) \cap V_{source} \neq \emptyset\}.
\end{align*}
The sub-hypergraph $G_{source} = (V_{source},E_{source})$ represents all template instances that contain at least one origin ID of interest and their hyperedges, as well as educts and products connected to their hyperedges for completeness. This sub-hypergraph captures all reachable labeled configurations originating from a given set of atom positions, providing a direct computational analogue to experimental label propagation.

A different perspective on $G$ focuses on a set of compound templates representing goal compounds, like measured products of a reaction network, for which the labeling should be investigated. Let $target$ be a set of position IDs that represent the compounds of interest. Let $V_{target} \subseteq V$ be a vertex set where
\begin{align*}
    V_{target} = \{v \in V \mid \{v_{p_1}, \dots, v_{p_{|v|}}\} \cap target \neq \emptyset\}
\end{align*}
and $E_{target} \subseteq E$ an edge set where
\begin{align*}
    E_{target} = \{(v_i, v_j) \mid~&v_i, v_j \in V_{target} \text{ and there exists a path } v_i \leadsto v_j \text{ in } G\\
    &\text{ that does not pass through any } v_k \in V_{target} \setminus \{v_i, v_j\}\}.
\end{align*}
The projected directed simple graph $G_{target} = (V_{target},E_{target})$ represents all template instances of the investigated goal compounds and the flow of labelings between them.

\subsection{Implementation Details}

Depending on the reaction network, the number of generated template instances can grow significantly. A memory-efficient data structure is utilized in the reference implementation to efficiently handle large quantities of instances. The implementation operates exclusively on template instances with origin-position pairs ordered by position ID. This enforced order allows instances of the same template to be stored in a hash map data structure where the keys represent an ordered tuple of template position IDs and the values represent a prefix tree (trie) of origin IDs. Each leaf of the prefix tree in combination with the key represents a complete template instances. This memory efficiency comes at the cost of runtime performance for insertion and iteration, compared to a normal set data structure.

To omit duplicate calculation of reactions with the same list of educts, a set of unvisited template instances is managed, reducing the number of educt combinations each step to those containing at least one unvisited template instance. This significantly increases runtime performance depending on the size of reaction network.


\section{Examples}\label{sec3}

Following, we explore examples with the saturation based method to demonstrate feasibility and usefulness.

\subsection{Tricarboxylic Acid Cycle}

\begin{figure}[H]
    \centering
    \includegraphics[width=\textwidth]{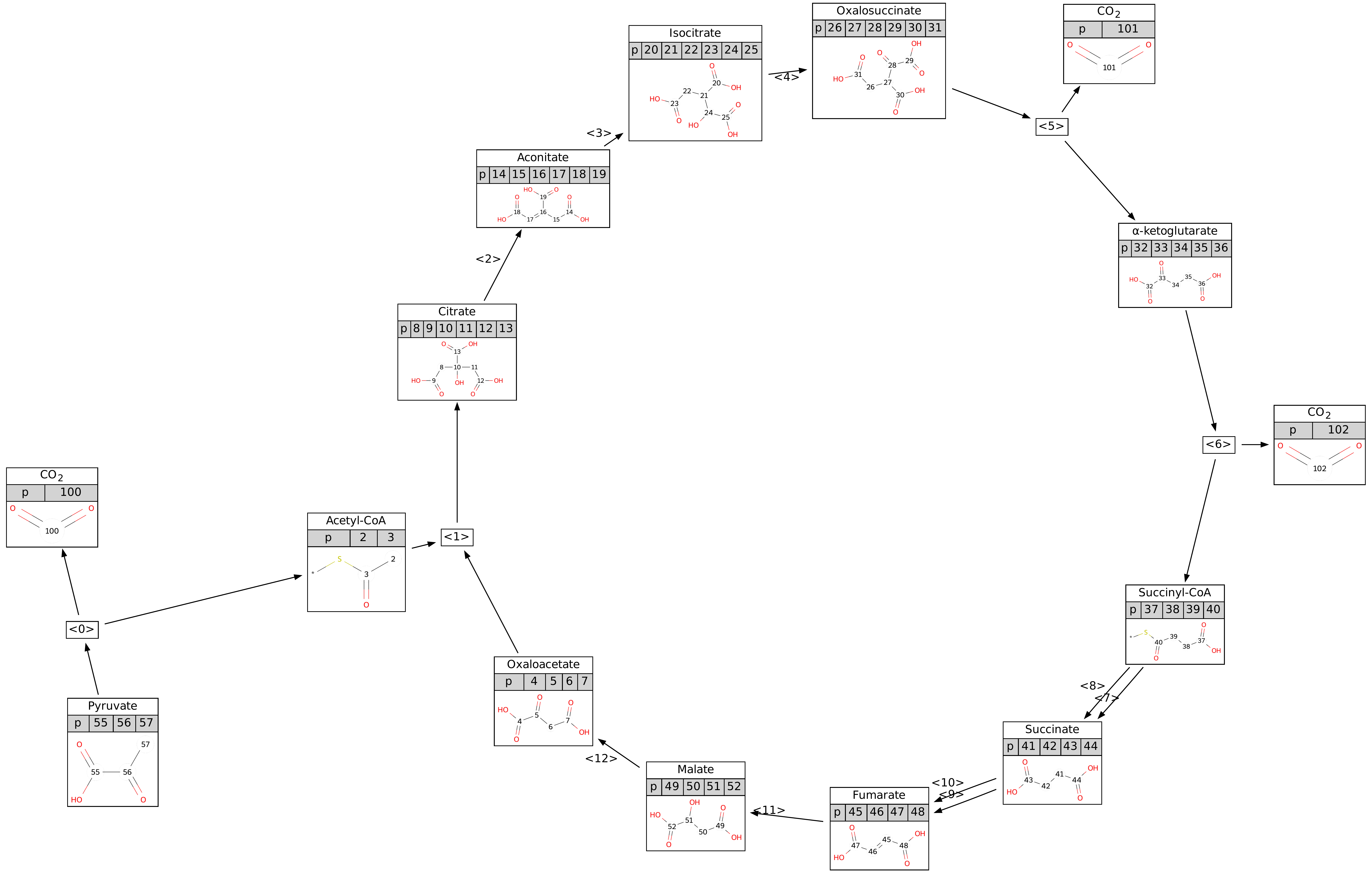}
    \caption{Hypergraph of a simple TCA cycle chemical reaction network CRN. Carbon atoms are uniquely numbered as visualized for each compound in the position ID row and molecular structure image. $CO_2$ is duplicated into three distinct compounds with different carbon position IDs for easier visualization and analysis. Compounds without carbon atoms are omitted.}
    \label{fig:tca_rn}
\end{figure}

The citric acid cycle, also known as the tricarboxylic acid cycle (TCA) cycle or the Krebs cycle, is a core and well-studied metabolic pathway in many different organisms. Atom tracing experiments have been performed in the past to elucidate the fate of individual carbons in the TCA cycle \emph{in vivo} as well as \emph{in silico}. \cite{Bruntz2017} review stable isotope-resolved metabolomics (SIRM) experiments for the TCA cycle. Dong \etal aggregated and analyzed results of $^{13}C$ and $^{2}H$ isotope tracing experiments \citep{Dong2019, MIT130611}. The semigroup approach for atom tracing by \cite{Njgaard2021} analyzed labeling patterns for the TCA cycle \emph{in silico}. Following, the TCA cycle is analyzed using the saturation-based method and compared to the results of the aforementioned studies.

\begin{figure}[H]
    \centering
    \includegraphics[width=\textwidth]{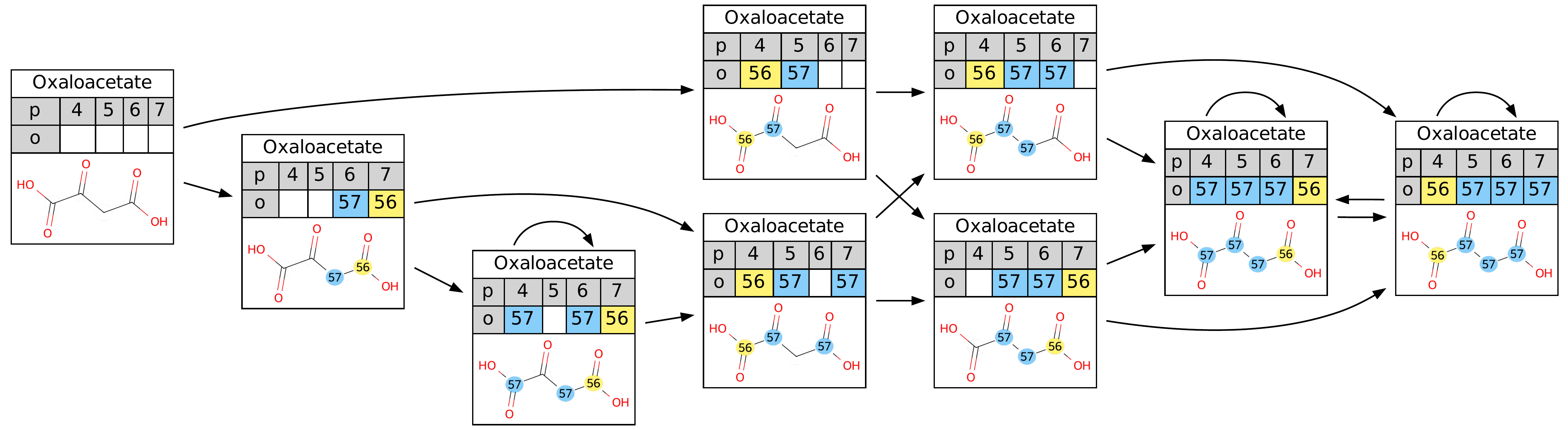}
    \caption{Oxaloacetate (OAA) target graph $G_{target}$ showing pyruvate carbon atom distribution in OAA from one TCA cycle to the next. Starting from unlabeled OAA, pyruvate carbon atoms accumulate until a steady configuration of OAA carbon provenance is reached. These automatically derived labeling transitions reproduce the experimentally validated carbon provenance patterns reported by Dong \etal.}
    \label{fig:tca_mod_pyruvate_oaa_cycles_trace}
\end{figure}

The chemical reaction network CRN for the TCA cycle is shown in \autoref{fig:tca_rn}. Compounds without carbon atoms are omitted. It resembles the same network as was used for the stereo-specific TCA example of Nøjgaard \etal, with the addition of a reaction from pyruvate to acetyl-CoA to match the same starting point as both other studies. For comparison with Figure 1 of Dong \etal, we defined label sets for all pyruvate carbon atoms to track their appearance in oxaloacetate (OAA) for each cycle of TCA. Furthermore, this analysis assumes a full saturation of the system with labeled pyruvate which can be reproduced by reducing the initial configuration $X_0$ to only a labeled pyruvate and an unlabled OAA:
\begin{align*}
    X_0 = \Big\{\big((55, 55), (56, 56), (57, 57)\big), \big((\bigcirc, 4), (\bigcirc, 5), (\bigcirc, 6), (\bigcirc, 7)\big)\Big\}
\end{align*}
The resulting hypergraph consists of 89 nodes and 93 hyperedges. To analyze and compare pyruvate carbon atoms in OAA, the graph $G_{target}$ for OAA is created as visualized in \autoref{fig:tca_mod_pyruvate_oaa_cycles_trace}. Starting from unlabeled OAA, each TCA cycle accumulates pyruvate carbon atoms in OAA until a steady configuration of OAA carbon provenance is reached in the strongly connected component of $\big((57, 4), (57, 5), (57, 6), (56, 7)\big)$ and $\big((56, 4), (57, 5), (57, 6), (57, 7)\big)$. These automatically derived labeling transitions reproduce the experimentally validated carbon provenance patterns reported by Dong \etal.

\begin{figure}[H]
    \centering
    \includegraphics[width=\textwidth]{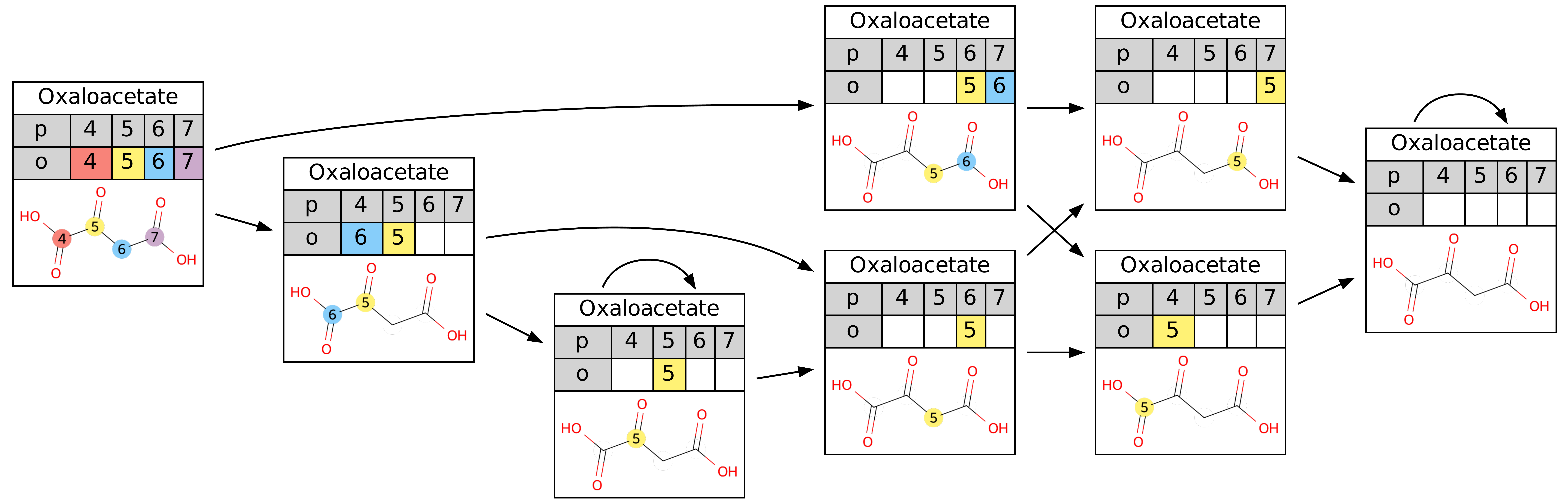}
    \caption{Oxaloacetate (OAA) target graph $G_{target}$ showing OAA carbon atom distribution in OAA from one TCA cycle to the next. Starting from fully labeled OAA, carbon atoms are continuously lost in the form of $CO_2$ until all have been replaced by carbon atoms originating from pyruvate. These automatically derived labeling transitions reproduce the simplified projected Cayley graph constructed by Nøjgaard \etal.}
    \label{fig:tca_mod_oaa_oaa_trace}
\end{figure}

Equally of interest is the fate of OAA carbon atoms through each TCA cycle. We used the normal initial configuration $X_0$ with label sets for the carbon atoms of OAA to generate the target graph $G_{target}$ of OAA. The resulting graph is visualized in \autoref{fig:tca_mod_oaa_oaa_trace}. Starting from a fully labeled OAA, carbon atoms are continuously lost in the form of $CO_2$ during the TCA cycle and replaced by carbon atoms from pyruvate. This process continues until OAA is fully unlabeled. Nøjgaard \etal constructed a simplified projected Cayley graph to represent this process, which matches the result of our method.

\subsection{EMP and ED Glycolysis}
\begin{figure}[H]
    \centering
    \includegraphics[width=\textwidth]{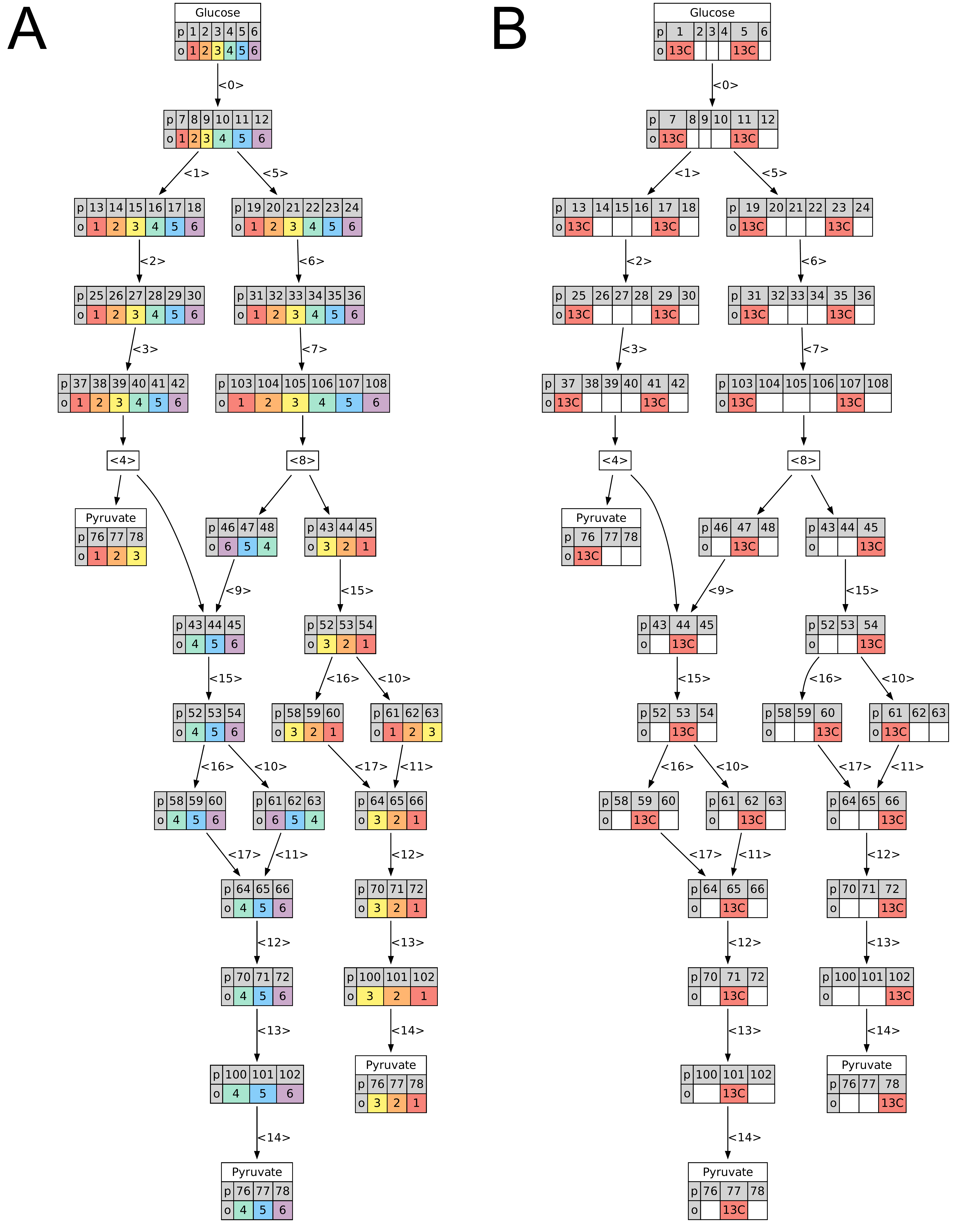}
    \caption{Pruned $G_{source}$ sub-hypergraph of the two Embden-Meyerhof-Parnas (EMP) and Entner-Doudoroff (ED) glycolysis pathways tracing glucose carbon atoms to pyruvate products. These automatically derived labeling paths reproduce the results of Andersen \etal.}
    \label{fig:emp_ed_graph}
\end{figure}

Atom tracing can be used to differentiate different pathways producing similar products. Glycolysis is a standard example where isotopic labeling allows differentiating two different types, the Embden-Meyerhof-Parnas (EMP) and Entner-Doudoroff (ED) glycolysis pathways. Both produce two pyruvate from a single glucose molecule \citep{Entner1952, Borodina2005}. Carbon tracing has shown that both pathways produce the same labeling in one pyruvate, but different labelings in the second \cite{Borodina2005}. Using the same graph grammar rules as \cite{Andersen2014}, a CRN of 25 vertices and 19 hyperedges was generated. We extracted atom-atom maps and used them to run a full labeling saturation. The constructed template instance hypergraph consists of 185 vertices and 176 hyperedges. As the glucose is the only carbon source in this scenario, the initial configuration $X_0$ can be modified as described above to only consider the atom position IDs of the glucose. This reduces the generated hypergraph to 42 vertices and 43 hyperedges. The sub-hypergraph $G_{source}$ was constructed and pruned to include only fully labeled template instances as visualized in \autoref{fig:emp_ed_graph}A.

Both pathways, including their shared intermediates and the distinct labeling of the second pyruvate, were reconstructed with complete atom provenance from the single carbon source glucose. Because no synthesis reactions are used, all resulting pyruvate template instances have full glucose atom provenance except for the pyruvate identity template:
\begin{align*}
    \big((76, 76), (77, 77), (78, 78)\big), \big((1, 76), (2, 77), (3, 78)\big), \big((3, 76), (2, 77), (1, 78)\big), \big((4, 76), (5, 77), (6, 78)\big)
\end{align*}
To plan and execute a lab experiment the minimal, easiest, or cheapest labeling must be found that is able to answer a specific research question. In this example, the minimal labeling that discerns all three pyruvate and, therefore, whether both pathways are active should be found. Using specific labeling sets for $X_0$ as described before allows us to simulate the labeling of two carbons in glucose with ${}^{13}C$ isotopes at atom position IDs $1$ and $5$. The resulting sub-hypergraph visualized in \autoref{fig:emp_ed_graph}B shows that the goal of discerning all three pyruvate and both pathways can be achieved when measuring the labeled positions in pyruvate.

\section{Discussion}\label{sec4}

Compared to isotopomer networks, EMUs, and similar, the proposed saturation-based method is a generalization in allowing tracing of an arbitrary number of labels instead of binary labelings. Furthermore, it relies solely on atom-atom maps without the need for experimental data or flux analysis. Isotopomer networks produce $2^N$ labelings for compounds with $N$ labeled atoms if no optimizations are applied. Given optimized label sets and initial configurations as described above or post-processing of the generated hypergraph $G$, the saturation-based approach produces similar binary labelings while reducing the configuration space to feasible traces only. Utilizing the full set of labels for $X_0$ can, under certain network criteria, lead to a larger number of template instances than isotopomer networks as the level of detail is significantly increased. The worst-case number of template instances is $|\Omega|^N$ for a compound with $N$ labeled atoms.

\begin{figure}[H]
    \centering
    \includegraphics[width=0.9\textwidth]{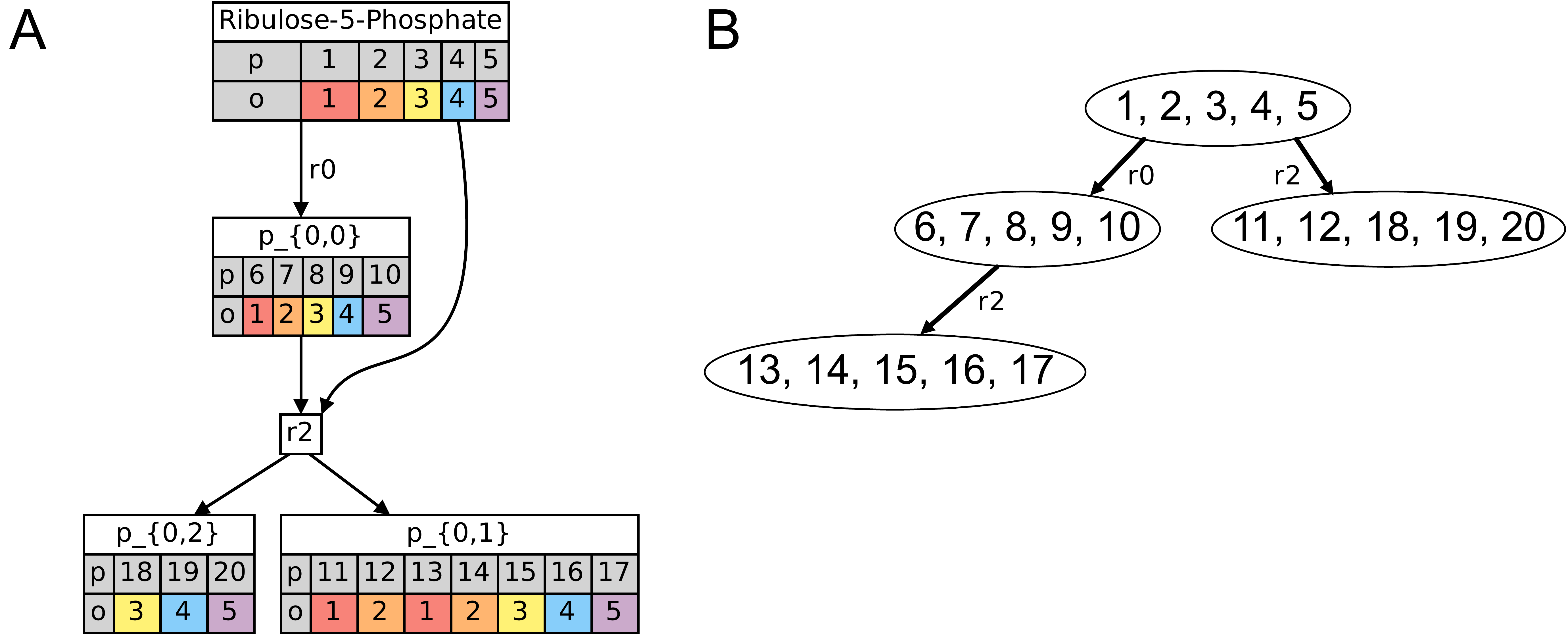}
    \caption{Comparison of saturation-based and semigroup-based atom tracing. \textbf{A}: Hypergraph representing two reactions, both using Ribulose-5-Phosphate, resulting in full atom provenance for reaction products. \textbf{B}: Projected Cayley graph of a semigroup-based tracing for the same two reactions, resulting only in partial atom provenance for reaction products.}
    \label{fig:multiplicity_vs_semigroups}
\end{figure}

Unlike previous frameworks, including the semigroup approach, this approach natively handles multiplicities, reuse, and synchronized multi-educt transformations, capabilities not simultaneously supported elsewhere. The difference in representing multiplicities and synchronicity compared to the semigroups approach is visualized in \autoref{fig:multiplicity_vs_semigroups}. As a generalization of the semigroups approach, the generated hypergraph $G$ can be projected to the Cayley graphs of the semigroup. This can be realized by tracking a specific set of labeled template instances via the atom-atom mappings of reaction hyperedges, eliminating multiplicities, reuse, and synchronicity.

Controlling the initial configuration $X_0$ of the saturation process enables fully automatic generation of labeling patterns for cyclical metabolic pathways such as TCA. In contrast to the manual labor of tracing atoms such as in Dong \etal, our new method significantly reduces the time to generate all patterns, is exact and not error prone, and produces a fully traceable hypergraph of entire chemical reaction networks.

Finally, the categorical formulation is not restricted to chemistry and may extend to any domain with well-defined compositional mappings. As long as exact mappings are provided on a set of IDs this approach can be applied on a generic level. Templates and template instances in that case represent different entities with some kind of fixed positions that provide information to subsequent entities via mappings.


\section{Conclusion}\label{sec5}

We have presented a novel saturation-based method for exhaustive enumeration of isotopomers in chemical reaction networks. By formalizing atom tracing as a Kleisli composition over the powerset monad, our framework enables compositional propagation of atom provenance while capturing multiplicities, reuse, and synchronized reaction semantics—features absent from existing isotopomer, EMU, or semigroup-based approaches.

The examples of formaldehyde condensation, the TCA cycle, and glycolysis illustrate that the approach reproduces experimentally established isotopic patterns and provides exact, fully traceable hypergraphs of atom flow. Where previous works are constrained by a limited number of labels or required additional data, the saturation-based method is fully independent, only requires atom-atom maps, and allows arbitrary numbers of labels. The detail of atom provenance that is fully automatically generated has previously been possible by manual tracing and reviewing as demonstrated by the MIT thesis and publications of Dong \etal. Customization of the initial configuration $X_0$ with specific label sets allows the immediate comparison of observed labelings, providing a direct applicability for experiment planning and metabolic network elucidation. Furthermore, the fully traceable hypergraph allows unbiased analysis of labeling patterns, otherwise introduced by flux balances and assumed metabolic steady configurations. For the near future, labeling experiments in \textit{E. coli} are already in the planning phase, informed by labeling patterns recommended by the saturation-based method.

Beyond chemistry, the categorical formalism underlying the saturation process applies to any system with discrete entities and compositional mappings, opening avenues for broader applications in network dynamics, provenance analysis, and formal systems modeling.


\section{Appendix}\label{sec:appendix}
Following, an extended formal definition of the method is provided that is closer to standard definitions of graph theory and graph grammars in chemistry. Note that for implementation the molecular structure of compounds is not relevant and only provided for relation to common methods.

Chemical compounds are defined as an undirected connected labeled graph $G=(V, E, l_V, l_E, p)$ consisting of a set $V(G)=V$ of vertices for the atoms and a set $E(G)=E$ of undirected edges for the bonds. Vertices and edges are equipped with attribute functions respectively. $l_V : V \rightarrow L_V$ associates each atom vertex with a value from the set $L_V$ representing atom types. $l_E : E \rightarrow L_E$ associates each edge with a value from the set $L_E$ representing bond types. Each atom vertex is further labeled with a unique position ID by the function $p : V(G) \rightarrow \mathbb{Z}$. The set of position IDs of a graph $G$ is defined as $p_G = \cup_{v \in V(G)} p(v)$. Position IDs uniquely identify vertices within a graph $|p_G| = |V(G)|$ as well as between graphs
\begin{align*}
    \forall (G, H), G \ne H: p_G \cap p_H = \emptyset.
\end{align*}

Given two compound graphs $G$ and $H$, a bijection $\varphi : V(G) \rightarrow V(H)$ is an isomorphism iff $\varphi$ is label-, attribute-, and edge-preserving:
\begin{align*}
    \forall v \in V(G) :&\ l_V(v) = l_V(\varphi(v))\\
    \forall v \in V(G) :&\ p(v) = p(\varphi(v))\\
    \forall (v, u) \in E(G) :&\ (\varphi(v), \varphi(u)) \in E(H)\\
    \forall (v, u) \in E(G) :&\ l_E(v, u) = l_E(\varphi(v), \varphi(u))
\end{align*}

Following the definition by \cite{Flamm2025}, a directed (multi)hypergraph $\mathscr{H}=(V, E)$ consists of a set $V(\mathscr{H})=V$ of vertices and a (multi)set $E(\mathscr{H})=E$ of directed hyperedges. To accommodate for multiplicities, hyperedges are of the form $(E^-, E^+)$ where $E^-$ and $E^+$ are multisets comprising elements of $V$. A chemical reaction network CRN is represented as a directed hypergraph where each vertex of the CRN is a connected graph representing a compound and each hyperedge is a chemical reaction. For each reaction $(E^-, E^+)$, $E^-$ and $E^+$ denote the educts and products of the reaction respectively. Stoichiometric coefficients of reactions are represented by the number of occurrence of compounds in $E^-$ and $E^+$.

An atom-atom map $AAM$ of a reaction $(E^-, E^+) \in E(\text{CRN})$ is a bijection
\begin{align*}
    \alpha : \biguplus\limits_{g \in E^-} V(g) \longrightarrow \biguplus\limits_{g \in E^+} V(g)
\end{align*}
from atoms of all compound graphs in $E^-$ to atoms of all compound graphs in $E^+$. Let $f : V(G) \rightarrow V'(G)$ be a function that selects a (sub)set of vertices $f(V(G)) \subseteq V(G)$ under an arbitrary condition. In chemistry, such a condition could be the selection of carbon atoms only. A partial atom-atom map is then a bijection
\begin{align*}
    \alpha_f : \biguplus\limits_{g \in E^-} f(V(g)) \longrightarrow \biguplus\limits_{g \in E^+} f(V(g))
\end{align*}
from a subset of atoms of all compound graphs in $E^-$ to a subset of atoms of all compound graphs in $E^+$.

For our approach, the notion of a compound graph is extended to a compound graph with origin $G^O=(V, E, l_V, l_E, p, o)$, adding a partial attribute function $o : f(V(G^O)) \rightarrow \mathbb{Z}$ that assigns origin IDs to vertices. This function is partial as only the atom vertices filtered by $f$ are assigned origin IDs. In contrast to position IDs, origin IDs are not unique and denote where the vertex, representing an atom, originated from before traversing reactions of the CRN. Given two compound graphs with origin $G^O$ and $H^O$, a bijection $\varphi^O : V(G^O) \rightarrow V(H^O)$ is an isomorphism iff $\varphi^O$ fulfills the same conditions as $\varphi$ as well as $\forall v \in V(G^O) : o(v) = o(\varphi^O(v))$.

Let a \emph{configuration} $X$ be a finite set of compound graphs with origin and $\mathscr{S} = (X, E, l_E)$ a directed multihypergraph. The attribute function $l_E : E(\mathscr{S}) \rightarrow AAM$ assigns each hyperedge an atom-atom map associated with a reaction of the chemical network.

An \emph{educt combination} of a reaction $c : E^- \rightarrow X$ assigns each educt a compound graph with origin of configuration $X$ allowing reuse and multiplicities, where $e \in E^-,~c(e) \in \{x \in X \mid e=\varphi(x)\}$ and $v \in V(e),~c(v) \in V(c(e))$. Note that for a single compound there may exist multiple compound graphs with origin, leading to the combinatorics of educt selection. Following, this combinatorial selection is referred to as an educt combination. The application of a reaction and corresponding atom-atom map on an educt combination $c$ generates a set of compound graphs with origin $E^+ \rightarrow G^O$ and, therefore, a new configuration $X'$. For this, origin IDs are mapped from educt to product vertices as
\begin{align*}
    g \in E^+,~v \in f(V(g)),~o(v) = o(c(\alpha_f^{-1}(v))).
\end{align*}

Let the continuous application of reactions and corresponding atom-atom maps on an initial configuration $X_0$ be a saturation process. This process is finite and closed once no new compound graphs are generated, e.g. $X_i=X_{i+1}$, and can be described in terms of a Kleisli category over a powerset monad. Each step of the saturation process updates the multihypergraph $\mathscr{S}$ to $V(\mathscr{S}) = X_{i+1}$ and by adding a hyperedge for each educt combination and reaction application to $E(\mathscr{S})$.

The powerset monad is defined as the triple $(\mathcal{P}_{fin}, \eta, \mu)$ consisting of the endofunctor $\mathcal{P}_{fin} : \mathbf{Set} \rightarrow \mathbf{Set}$ and the two natural transformations $\eta$ and $\mu$. The unit $\eta$ is defined as $\eta_Y : Y \rightarrow \mathcal{P}_{fin}(Y)$, assigning to an element of $Y$ the set consisting of that single element $\eta_Y(y) = \{y\}$. The multiplication $\mu$ flattens nested finite sets into a single set through union: $\mu_Y : \mathcal{P}_{fin}(\mathcal{P}_{fin}(Y)) \rightarrow \mathcal{P}_{fin}(Y)$, $\mu_Y(y) = \bigcup y$.

Further, the Kleisli category $C_{\mathcal{P}_{fin}}$ for the powerset monad is defined, with sets as objects and morphisms $f : Y \rightarrow Z$ of the form $Y \rightarrow \mathcal{P}_{fin}(Z)$. Given two morphisms $f : Y \rightarrow \mathcal{P}_{fin}(Z)$ and $g : Z \rightarrow \mathcal{P}_{fin}(W)$ their Kleisli composition is defined as $g \odot f = \mu_W \circ \mathcal{P}_{fin}(g) \circ f$.

Given a configuration $X$ the reaction application morphism \emph{step} is defined as $step : \mathcal{P}_{fin}(Y) \rightarrow \mathcal{P}_{fin}(\mathcal{P}_{fin}(Y))$ in $C_{\mathcal{P}_{fin}}$ which applies all reactions and corresponding atom-atom maps $R \in \mathcal{R}$ on all educt combinations $c_R$ of $X^*$ and produces a set of new configurations by:
\begin{align*}
    step(X) = \big\{ X \cup c_r(E^-) \mid r=(E^-,E^+) \in E(\text{CRN}),~\alpha_f \in r,~c_r \in X^* \big\}
\end{align*}
Because $E(\text{CRN})$ and $c_r$ are both finite, $step(X)$ is a finite set of successor configurations. Flattening via $\mu$, $\mu_Y \circ step : \mathcal{P}_{fin}(Y) \rightarrow \mathcal{P}_{fin}(Y)$ can be viewed as a Kleisli morphism $step : Y \rightarrow_{C_{\mathcal{P}_{fin}}} Y$, representing a single transition step of the system.

The process of applying reactions to a configuration involves iterating through educt combinations and updating the configuration accordingly. Iteratively applying reactions expands the configuration with new compound graphs with origin until a fixed point is reached where no further graphs appear. The n-fold Kleisli composition is expressed as
\begin{align*}
    step^{(n)} = \underbrace{step \odot \cdots \odot step}_{n~\text{times}},
\end{align*}
representing all configurations reachable in $n$ steps. A saturation process can be defined by choosing $n$ so that the produced configuration is closed
\begin{align*}
    step^* = \bigcup\limits_{n \geq 0} step^{(n)},
\end{align*}
representing all reachable configurations under repeated application of all reactions $E(\text{CRN})$.


\section*{Funding}

This work was supported by Novo Nordisk Foundation grant "Mathematical Modelling for Microbial Community Induced Metabolic Diseases" (MATOMIC) with Grant Reference: NNF21OC0066551.

\bibliographystyle{unsrtnat}
\bibliography{sample}

@article{Njgaard2021,
  title     = {Cayley Graphs of Semigroups Applied to Atom Tracking in Chemistry},
  volume    = {28},
  issn      = {1557-8666},
  doi       = {10.1089/cmb.2020.0548},
  number    = {7},
  journal   = {Journal of Computational Biology},
  publisher = {Mary Ann Liebert Inc},
  author    = {Nøjgaard,  Nikolai and Fontana,  Walter and Hellmuth,  Marc and Merkle,  Daniel},
  year      = {2021},
  month     = jul,
  pages     = {701–715}
}

@article{Schmidt1997,
  title     = {Modeling isotopomer distributions in biochemical networks using isotopomer mapping matrices},
  volume    = {55},
  issn      = {1097-0290},
  doi       = {10.1002/(sici)1097-0290(19970920)55:6<831::aid-bit2>3.0.co;2-h},
  number    = {6},
  journal   = {Biotechnology and Bioengineering},
  publisher = {Wiley},
  author    = {Schmidt,  Karsten and Carlsen,  Morten and Nielsen,  Jens and Villadsen,  John},
  year      = {1997},
  month     = sep,
  pages     = {831–840}
}

@article{Wiechert1999,
  title     = {Bidirectional reaction steps in metabolic networks: III. Explicit solution and analysis of isotopomer labeling systems},
  volume    = {66},
  issn      = {1097-0290},
  doi       = {10.1002/(sici)1097-0290(1999)66:2<69::aid-bit1>3.0.co;2-6},
  number    = {2},
  journal   = {Biotechnology and Bioengineering},
  publisher = {Wiley},
  author    = {Wiechert,  Wolfgang and Möllney,  Michael and Isermann,  Nichole and Wurzel,  Michael and de Graaf,  Albert A.},
  year      = {1999},
  pages     = {69–85}
}

@article{Antoniewicz2007,
  title     = {Elementary metabolite units (EMU): A novel framework for modeling isotopic distributions},
  volume    = {9},
  issn      = {1096-7176},
  doi       = {10.1016/j.ymben.2006.09.001},
  number    = {1},
  journal   = {Metabolic Engineering},
  publisher = {Elsevier BV},
  author    = {Antoniewicz,  Maciek R. and Kelleher,  Joanne K. and Stephanopoulos,  Gregory},
  year      = {2007},
  month     = jan,
  pages     = {68–86}
}

@article{RMG3,
  author    = {Liu, Mengjie and Grinberg Dana, Alon and Johnson, Matthew S. and Goldman, Mark J. and Jocher, Agnes and Payne, A. Mark and Grambow, Colin A. and Han, Kehang and Yee, Nathan W. and Mazeau, Emily J. and Blondal, Katrin and West, Richard H. and Goldsmith, C. Franklin and Green, William H.},
  title     = {{Reaction Mechanism Generator v3.0: Advances in Automatic Mechanism Generation}},
  journal   = {Journal of Chemical Information and Modeling},
  volume    = {61},
  number    = {6},
  pages     = {2686-2696},
  year      = {2021},
  month     = may,
  doi       = {10.1021/acs.jcim.0c01480},
}

@inbook{Andersen2016,
  title     = {A Software Package for Chemically Inspired Graph Transformation},
  isbn      = {9783319405308},
  issn      = {1611-3349},
  doi       = {10.1007/978-3-319-40530-8_5},
  booktitle = {Graph Transformation},
  publisher = {Springer International Publishing},
  author    = {Andersen,  Jakob L. and Flamm,  Christoph and Merkle,  Daniel and Stadler,  Peter F.},
  year      = {2016},
  pages     = {73–88}
}

@article{Jang2018,
  title     = {Metabolomics and Isotope Tracing},
  volume    = {173},
  issn      = {0092-8674},
  doi       = {10.1016/j.cell.2018.03.055},
  number    = {4},
  journal   = {Cell},
  publisher = {Elsevier BV},
  author    = {Jang,  Cholsoon and Chen,  Li and Rabinowitz,  Joshua D.},
  year      = {2018},
  month     = may,
  pages     = {822–837}
}

@article{Golnik2025,
  title     = {Atom Transition Networks and Isotope Labeling Patterns in Large Chemical Reaction Networks},
  doi       = {10.21203/rs.3.rs-5888287/v1},
  publisher = {Springer Science and Business Media LLC},
  author    = {Golnik,  Richard and Stadler,  Peter Florian and Gatter,  Thomas},
  year      = {2025},
  month     = jan 
}

@inbook{Andersen2014,
  title     = {50 Shades of Rule Composition},
  ISBN      = {9783319103983},
  ISSN      = {1611-3349},
  url       = {http://dx.doi.org/10.1007/978-3-319-10398-3_9},
  DOI       = {10.1007/978-3-319-10398-3_9},
  booktitle = {Formal Methods in Macro-Biology},
  publisher = {Springer International Publishing},
  author    = {Andersen,  Jakob Lykke and Flamm,  Christoph and Merkle,  Daniel and Stadler,  Peter F.},
  year      = {2014},
  pages     = {117–135}
}

@article{Dong2019,
  title     = {{Dissecting Mammalian Cell Metabolism through $^{13}$C- and $^{2}$H-Isotope Tracing: Interpretations at the Molecular and Systems Levels}},
  volume    = {59},
  ISSN      = {1520-5045},
  url       = {http://dx.doi.org/10.1021/acs.iecr.9b05154},
  DOI       = {10.1021/acs.iecr.9b05154},
  number    = {6},
  journal   = {Industrial \& Engineering Chemistry Research},
  publisher = {American Chemical Society (ACS)},
  author    = {Dong,  Wentao and Moon,  Sun Jin and Kelleher,  Joanne K. and Stephanopoulos,  Gregory},
  year      = {2019},
  month     = nov,
  pages     = {2593–2610}
}

@phdthesis{MIT130611,
  author    = {Dong, Wentao},
  title     = {Exploring Cancer Metabolism through Isotopic Tracing and Metabolic Flux Analysis},
  school    = {Massachusetts Institute of Technology},
  year      = {2020},
  type      = {PhD thesis},
  address   = {Cambridge, MA, USA},
  url       = {https://dspace.mit.edu/handle/1721.1/130611},
}

@article{Bruntz2017,
  title     = {Exploring cancer metabolism using stable isotope-resolved metabolomics (SIRM)},
  volume    = {292},
  ISSN      = {0021-9258},
  url       = {http://dx.doi.org/10.1074/jbc.R117.776054},
  DOI       = {10.1074/jbc.r117.776054},
  number    = {28},
  journal   = {Journal of Biological Chemistry},
  publisher = {Elsevier BV},
  author    = {Bruntz,  Ronald C. and Lane,  Andrew N. and Higashi,  Richard M. and Fan,  Teresa W.-M.},
  year      = {2017},
  month     = jul,
  pages     = {11601–11609}
}

@article{deFalco2022,
  title     = {Metabolic flux analysis: a comprehensive review on sample preparation,  analytical techniques,  data analysis,  computational modelling,  and main application areas},
  volume    = {12},
  ISSN      = {2046-2069},
  url       = {http://dx.doi.org/10.1039/D2RA03326G},
  DOI       = {10.1039/d2ra03326g},
  number    = {39},
  journal   = {RSC Advances},
  publisher = {Royal Society of Chemistry (RSC)},
  author    = {de Falco,  Bruna and Giannino,  Francesco and Carteni,  Fabrizio and Mazzoleni,  Stefano and Kim,  Dong-Hyun},
  year      = {2022},
  pages     = {25528–25548}
}

@article{Bey2021,
  title     = {Robustifying Experimental Tracer Design for13C-Metabolic Flux Analysis},
  volume    = {9},
  ISSN      = {2296-4185},
  url       = {http://dx.doi.org/10.3389/fbioe.2021.685323},
  DOI       = {10.3389/fbioe.2021.685323},
  journal   = {Frontiers in Bioengineering and Biotechnology},
  publisher = {Frontiers Media SA},
  author    = {Beyß,  Martin and Parra-Peña,  Victor D. and Ramirez-Malule,  Howard and N\"{o}h,  Katharina},
  year      = {2021},
  month     = jun 
}

@article{Flamm2025,
  title     = {Assembly in Directed Hypergraphs},
  volume    = {481},
  ISSN      = {1471-2946},
  url       = {http://dx.doi.org/10.1098/rspa.2025.0331},
  DOI       = {10.1098/rspa.2025.0331},
  number    = {2324},
  journal   = {Proceedings of the Royal Society A: Mathematical,  Physical and Engineering Sciences},
  publisher = {The Royal Society},
  author    = {Flamm,  Christoph and Merkle,  Daniel and Stadler,  Peter F.},
  year      = {2025},
  month     = oct 
}

@article{Borodina2005,
  title     = {Metabolic Network Analysis of Streptomyces tenebrarius, a Streptomyces Species with an Active Entner-Doudoroff Pathway},
  volume    = {71},
  ISSN      = {1098-5336},
  url       = {http://dx.doi.org/10.1128/AEM.71.5.2294-2302.2005},
  DOI       = {10.1128/aem.71.5.2294-2302.2005},
  number    = {5},
  journal   = {Applied and Environmental Microbiology},
  publisher = {American Society for Microbiology},
  author    = {Borodina,  Irina and Schöller,  Charlotte and Eliasson,  Anna and Nielsen,  Jens},
  year      = {2005},
  month     = may,
  pages     = {2294–2302}
}

@article{Entner1952,
  title     = {GLUCOSE AND GLUCONIC ACID OXIDATION OF PSEUDOMONAS SACCHAROPHILA},
  volume    = {196},
  ISSN      = {0021-9258},
  url       = {http://dx.doi.org/10.1016/S0021-9258(19)52415-2},
  DOI       = {10.1016/s0021-9258(19)52415-2},
  number    = {2},
  journal   = {Journal of Biological Chemistry},
  publisher = {Elsevier BV},
  author    = {Entner,  Nathan and Doudoroff,  Michael},
  year      = {1952},
  month     = jun,
  pages     = {853–862}
}

\end{document}